# Semiconductor quantum computer design with 100 nm separation of nuclear-spin qubits

**Vladimir Privman and Dima Mozyrsky**

*Department of Physics and Center for Advanced Materials Processing, Clarkson University, Potsdam, NY 13699-5820, USA*
*Telephone (315) 268-3891, fax(315) 268-6610, e-mail address privman@clarkson.edu*

**Abstract:** We combine elements of the 1998 quantum computing proposals by Privman, Vagner and Kventsel, and by Kane, with the new idea of nuclear-spin qubit interactions mediated indirectly via the bound outer electrons of impurity atoms whose nuclear spins 1/2 are the qubits. These electrons, in turn, interact via the two-dimensional electron gas in the quantum Hall effect regime. The resulting quantum computing scheme retains all the gate-control and measurement aspects of the proposal by Kane, but allows qubit spacing at distances of order 100 nm, attainable with the present-day semiconductor-heterostructure device technologies.

## 1. Introduction

The general layout of a solid-state quantum computer is shown in Figure 1. Qubits are positioned with precision of few nanometers in a heterostructure. One must propose how to effect and control single-qubit interactions, two-qubit interactions, and explore how the controlled dynamics owing to these interactions compares to decoherence and relaxation. The proposal must include ideas for implementation of initialization, readout, and gate functions.

We outline results for models of quantum computing with nuclear spins as qubits, and with coupling mediated by the two-dimensional electron gas in the integer quantum Hall effect state [1-3]. In strong magnetic fields, the spatial states of the electrons confined in the two-dimensional layer in which the qubits are placed, are quantized by the field to have properties like free-space Landau levels. The lattice potential and the impurities actually cause formation of narrow bands instead of the sharp levels, separated by localized states. As a result, for ranges of magnetic field, the localized states fill up while the extended states resemble completely filled integer number of Landau levels. These states are further Zeeman split owing to the electron spin. At low temperatures, one can find field values such that only one Zeeman sublevel is completely filled in the ground state.

The electronic state in such systems, that show the quantum Hall effect in conductivity, are highly correlated and nondissipative. If nuclear spins are used as qubits, i.e., atoms with nuclear spin 1/2 are sparsely positioned in the zero-nuclear spin host, such as the zero-nuclear-spin isotope 28 of Si, which constitutes 92% of natural silicone, then their zero-temperature relaxation will be slowed down, $T_1 = O(10^3)$ sec.

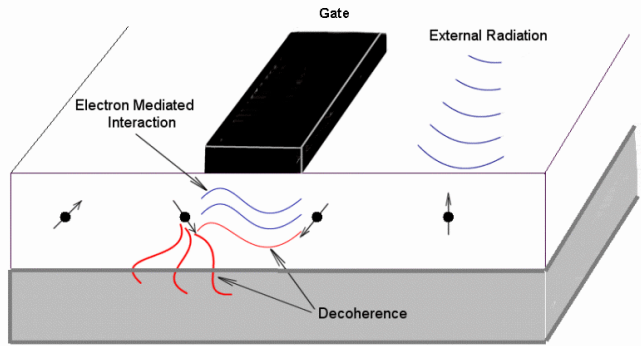

Fig. 1. Schematic illustration of a semiconductor heterostructure quantum information processor. The qubits, represented by the arrows overlaying heavy dots, are spins 1/2 of nuclei or localized electrons. Individual control of the temporal evolution of the spins can be achieved with the use of external electromagnetic radiation, i.e., NMR or ESR pulses. The spins are also coupled with each other via interaction mediated by the two-dimensional electron gas in the heterostructure, or by other means. The external and internal interactions can be controlled by gates formed on top of the heterostructure. The external environment, that includes crystal lattice, electron gas, defects, impurity potentials, causes relaxation and decoherence of the qubits.

Localized spins, both nuclear and electronic, interact by exchanges of spin excitons—spin waves consisting of a superposition of bound electron-hole pair states. The spectrum of these excitations [4,5], observed experimentally in [6], has a gap corresponding to the Zeeman splitting. This gap is the cause of slow relaxation and decoherence. The exchange of virtual spin excitons mediates the qubit-qubit interaction and also, via scattering of virtual excitons from impurity potentials, relaxation and decoherence of single qubits.

The original proposal to use nuclear spin qubits directly coupled by the two-dimensional electron gas [3], required positioning the qubits at distances comparable to several magnetic lengths. The latter is of order 10 nm for magnetic fields of several Tesla. The qubit-qubit interaction decays exponentially on this length scale. Recently, we proposed a new improved model [2] in which the qubit interactions are mediated via coupling of the two-dimensional electron gas to the outer impurity electrons. This applies if the atoms, whose nuclear spins are the qubits, are single-electron donors such as the isotope 31 of P. These phosphorous impurities were originally utilized in the model of Kane [7] where they must be actually positioned at separations of about 4 nm for the wavefunctions of the outer electrons, which are bound at low temperatures, to overlap significantly.

## 2. The new improved model

In our new model [2], with nuclear spins coupling to the outer bound electrons which, in turn, interact via the two-dimensional electron gas, the interaction turned out to be of a much longer range as compared to the model of [7]: the qubit separation can be of order 100 nm. Another advantage is that gate control of the individual qubits and of qubit-qubit interactions is possible. We have carried out extensive perturbative many-body calculations [1-3,8] allowing estimation of $T_{int}$ and $T_2$ for both the original quantum-computing proposal [3] and its improved version [2], where the main improvement is in the possibility of the gate control along the lines of [7]. The "clock speed" of the improved model is also faster by about two orders of magnitude.

The results are summarized in Table 1. We show estimates of all four relevant time scales for the two models introduced earlier. The "original" model [3] corresponds to nuclear spins 1/2 introduced at qubits in atoms without an outer loosely bound electron. The "improved" model corresponds to the case when the outer electron is present and its interaction with the nuclear spin and the two-dimensional electron gas dominates the dynamics.

Table 1. Time scales of the qubit dynamics for the original [3] and improved [2] versions of the nuclear spin quantum computer with interactions mediated by the two-dimensional electron gas.

|  | The original model | The improved model | Definition of the time scale |
| --- | --- | --- | --- |
| $T_{ext}$ | $O(10^{-5})$ sec | $O(10^{-5})$ sec | Single-qubit external NMR-radiation control time |
| $T_{int}$ | $O(1)$ sec | $O(10^{-2})$ sec | Time scale defined by the two-qubit interactions |
| $T_1$ | $O(10^3)$ sec | $O(10)$ sec | Time scale associated with energy relaxation |
| $T_2$ | $O(10)$ sec | $O(10^{-1})$ sec | Intrinsic quantum-mechanical decoherence time |

The data shown in Table 1 were obtained assuming typical parameters for the standard heterojunctions utilized in quantum-Hall-effect experiments today, and qubit separation of 65 nm. Thus, the parameter values taken [2,3] were more appropriate for the GaAs system than for Si, even though the main isotopes of gallium and arsenic have nuclear spin 3/2 and cannot serve as spin-zero hosts. The reason for using these values has been that experimental verification of some of the numbers might be possible in the available materials before cleaner and different composition materials needed for quantum computing are produced.

Our estimates, see Table 1, indicate that the desired [9] quality factor $Q = T_{int} / T_2 = 10^{-5}$ is not obtained for the present system. Actually, no quantum computing proposal to date, scalable by other criteria, satisfies the $10^{-5}$ quality-factor criterion. The values range from $10^{-1}$ to $10^{-2}$. The resolution could come from development of better error-correction algorithms or from improving the physical system to obtain a better quality factor.

In our estimation of the decoherence time scale, we used parameters typical of a standard, "dirty" heterostructure with large spatial fluctuations of the impurity potential. These heterostructures have been suitable for standard experiments because they provide wider quantum-Hall plateaus, i.e., ranges of magnetic field for which all the extended states of a Zeeman sublevel are filled. Much cleaner, ultra-high mobility structures can be obtained by placing the ionized impurity layer at a larger distance from the two-dimensional gas or by injecting conduction electrons into the heterostructure by other means. Thus, our quantum-computing proposals [2,3] are unique not only in the large qubit separation allowed but also in that there is a clear direction of exploration to allow physical, rather than algorithmic, resolution of the quality factor problem. This possibility should be further explored both experimentally and theoretically.

The authors acknowledge useful discussions and collaboration with M. L. Glasser, R. G. Mani, L. S. Schulman and I. D. Vagner. This research was supported by the National Security Agency (NSA) and Advanced Research and Development Activity (ARDA) under Army Research Office (ARO) contract number DAAD 19-99-1-0342.